\documentclass{ws-ijmpd}

\newcommand{\eps}{\varepsilon}

\newcommand{\la}{\lambda}
\newcommand{\si}{\sigma}
\newcommand{\Ga}{\Gamma}
\newcommand{\eq}[2]{\begin{equation} #1 \label{#2} \end{equation}}

\newcommand{\beqs}{\begin{equation*}}
\newcommand{\beq}{\begin{equation}}
\newcommand{\eeqs}{\end{equation*}}
\newcommand{\eeq}{\end{equation}}

\newcommand{\cO}{{\cal O}}

\begin{document}


\markboth{D.~GRUMILLER AND N.~JOHANSSON}
{CONSISTENT BOUNDARY CONDITIONS FOR CTMG AT THE CHIRAL POINT}

\catchline{}{}{}{}{}

\title{CONSISTENT BOUNDARY CONDITIONS FOR COSMOLOGICAL TOPOLOGICALLY MASSIVE GRAVITY AT THE CHIRAL POINT}

\author{D.~GRUMILLER}

\address{
 Center for Theoretical Physics,
Massachusetts Institute of Technology,\\
77 Massachusetts Ave.,
Cambridge, MA  02139, USA\\ and \\
Institute for Theoretical Physics, Vienna University of Technology,\\
Wiedner Hauptstr.~8--10/136, Vienna, A-1040, Austria\\
grumil@lns.mit.edu
}

\author{N.~JOHANSSON}

\address{Institutionen f{\"o}r Fysik och Astronomi, Uppsala Universitet,\\
Box 803, S-751 08 Uppsala, Sweden\\
Niklas.Johansson@fysast.uu.se
}

\maketitle

\begin{history}
MIT-CTP 3972, UUITP-18/08, {\tt arXiv:0808.2575}

\received{1.~October 2008}
\accepted{2.~October 2008}
\comby{D.V.~Ahluwalia}
\end{history}

\begin{abstract}
We show that cosmological topologically massive gravity at the chiral point allows not only Brown--Henneaux boundary conditions as consistent boundary conditions, but slightly more general ones which encompass the logarithmic primary found in {{\tt 0805.2610}} as well as all its descendants. 
\end{abstract}



\section{Introduction}

Cosmological topologically massive gravity~\cite{Deser:1982sv} (CTMG) is a 3-dimensional theory of gravity that exhibits gravitons~\cite{Deser:1982vy,Deser:1982wh} and black holes~\cite{Banados:1992wn}. With the sign conventions of Ref.~\cite{Grumiller:2008qz} its action is given by
\begin{equation}
I_{\rm CTMG} 
= \frac{1}{16\pi G}\int d^3x\sqrt{-g}\,\Big[R+\frac{2}{\ell^2} 
+\frac{1}{2\mu} \,\eps^{\la\mu\nu}\,\Ga^\rho{}_{\la\si}\,\big(\partial_\mu \Ga^\si{}_{\nu\rho}+\frac23 \,\Ga^\si{}_{\mu\tau}\Ga^\tau{}_{\nu\rho}\big)\Big] \,,
\label{eq:cg1}
\end{equation}
where the negative cosmological constant is parameterized by $\Lambda=-1/\ell^2$. If the constants $\mu$ and $\ell$ satisfy the condition
\eq{
\mu\ell = 1
}{eq:cg2}
the theory is called ``CTMG at the chiral point''. The condition \eqref{eq:cg2} is special because one of the central charges of the dual boundary CFT vanishes, $c_L=0$, $c_R\neq 0$. 

This observation was the motivation for Ref.~\cite{Li:2008dq} to consider CTMG at the chiral point in some detail. In that work the theory \eqref{eq:cg1} with \eqref{eq:cg2} was dubbed ``chiral gravity'', assuming that all solutions obey the Brown--Henneaux boundary conditions~\cite{Brown:1986nw}. Moreover, it was conjectured that CTMG at the chiral point is a chiral theory and that the local physical degree of freedom, the topologically massive graviton, disappears. These statements were disputed in Ref.~\cite{Carlip:2008jk}, which engendered a lot of recent activity concerning CTMG~\cite{Hotta:2008yq,Grumiller:2008qz,
Carlip:2008qh,Giribet:2008bw}. In particular, the present authors constructed explicitly~\cite{Grumiller:2008qz} a physical mode not considered in Ref.~\cite{Li:2008dq} using their formalism. This mode, which we call ``logarithmic primary'', violates the Brown--Henneaux boundary conditions. These results were confirmed very recently~\cite{Giribet:2008bw}, where one of the descendants of
the logarithmic primary was considered. It was found that this descendant (and all successive descendants) can be made consistent with the Brown--Henneaux boundary conditions by a diffeomorphism. Thus, these modes are present in classical CTMG (in addition to the standard boundary gravitons), even if Brown--Henneaux boundary conditions are imposed. The latest development is a simple classical proof~\cite{Strominger:2008dp} of the chirality of the generators of diffeomorphisms at $\mu\ell=1$, concurrent with previous results~\cite{Carlip:2008qh}.

In the conclusions of Ref.~\cite{Strominger:2008dp} it was speculated that perhaps there are consistent boundary conditions other than the ones by Brown and Henneaux for CTMG at the chiral point. It is the purpose of this note to show that this is indeed the case and that the new set of boundary conditions encompasses the logarithmic primary.

\section{Beyond Brown--Henneaux}

We follow as closely as possible the notation and the logical flow of Ref.~\cite{Strominger:2008dp}. Any metric consistent with the boundary conditions to be imposed below must asymptote to AdS$_3$, which in Poincar\'e coordinates is given by
\eq{
g_{\mu\nu}^{\rm AdS} \,dx^\mu\,dx^\nu=\ell^2\,\left(\frac{dx^+\,dx^-+dy^2}{y^2}\right)\,,
}{eq:false1}  
where the boundary is located at $y=0$. The Brown--Henneaux boundary conditions then require that fluctuations $h_{\mu\nu}$ of the metric about \eqref{eq:false1} must fall off as
\eq{
\left(\begin{array}{ccc}
h_{++}={\cal O}(1) & h_{+-} = {\cal O} (1) & h_{+y} = {\cal O}(y) \\
                   & h_{--} = {\cal O} (1) & h_{-y} = {\cal O}(y) \\
                   &                       & h_{yy} = {\cal O}(1)
\end{array}\right).
}{eq:false2}
By $\cO(x)$ we mean that the corresponding fluctuation metric component behaves {\em at most} proportional to $x$ in the small $y$ limit.

We define now suitable boundary conditions that encompass the logarithmic primary and its descendants. 
Let us first recall the form of the
logarithmic primary and see how the Brown--Henneaux boundary conditions need to be weakened. Translating the result (3.3) of Ref.~\cite{Grumiller:2008qz} into Poincar\'e coordinates yields schematically\footnote{The coordinates in that work are related to the coordinates here as follows: $x^\pm=(\phi\mp\tau)/2$, $y \sim e^{-\rho}$.}
\eq{
h^{\rm new}_{\rm \mu\nu}\, dx^\mu\, dx^\nu \sim \cO (\log y) \,(dx^-)^2 + \cO (y\log y)\,dx^- \,dy + \cO (y^2\log y)\,dy^2\,.
}{eq:false5}
Evidently the logarithmic primary behaves as follows:
\beq
h^{\rm new}_{--} = \cO (\log y)\,,\quad h^{\rm new}_{-y} = \cO (y\log y)\,,\quad h^{\rm new}_{yy} = \cO (y^2\log y).
\label{eq:false6}
\eeq
From \eqref{eq:false2} we see that the Brown--Henneaux boundary conditions for these three components are
\beq
 h_{--} = \cO (1)\,, \qquad h_{-y} = \cO (y)\,, \qquad h_{yy} = \cO (1)\,.
\label{eq:false7}
\eeq
It is clear that \eqref{eq:false6} is incompatible with \eqref{eq:false7}. However, only the first two conditions of \eqref{eq:false1} have to be weakened logarithmically to encompass the logarithmic primary.

Therefore, we propose the following set of boundary conditions:\footnote{The proposal \eqref{BC} may be compared with footnote 3 in Ref.~\cite{Strominger:2008dp}: it is not necessary to weaken the boundary conditions of all components $h_{\pm\pm}$ to $\cO(\ln{y})$ (see first sentence) and it is not sufficient to take only $h_{--}$ to be $\cO(\ln{y})$ (see second sentence).}
\beq \label{BC}
\left( \begin{array}{lll}
h^{\rm new}_{++} = \cO(1) & h^{\rm new}_{+-} = \cO(1)    & h^{\rm new}_{+y} = \cO(y) \\
                  & h^{\rm new}_{--} = \cO(\log y) & h^{\rm new}_{-y} = \cO(y\log y) \\
                  &                      & h^{\rm new}_{yy} = \cO(1)
\end{array} \right)
\eeq
Let us determine the diffeomorphisms 
\eq{
g_{\mu\nu} = g_{\mu\nu}^{\rm AdS} + h^{\rm new}_{\mu\nu} \quad \to \quad {\cal L}_\zeta\, g_{\mu\nu} = \tilde g_{\mu\nu} = g_{\mu\nu}^{\rm AdS} + \tilde h^{\rm new}_{\mu\nu}
}{eq:false9}
that preserve these boundary conditions. I.e., we require that also $\tilde h^{\rm new}_{\mu\nu}$ has the fall-off behavior postulated in \eqref{BC}. 
Calculating the generator $\zeta^\mu$ with this requirement obtains
\begin{align}
\zeta^+ &= \epsilon^+(x^+)-\frac{y^2}{2} \,\partial_-^2\epsilon^- + \cO(y^4\log{y})\\
\zeta^- &= \epsilon^-(x^-)-\frac{y^2}{2} \,\partial_+^2\epsilon^+ + \cO(y^4)\\
\zeta^y &= \frac y2 \left(\partial_+\epsilon^+(x^+)+\partial_-\epsilon^-(x^-)\right) + \cO (y^3)
\end{align}
Remarkably, the only difference to the Brown--Henneaux-case is the possibility of an $\cO(y^4 \log y)$ behavior for the sub-sub-leading terms in the $\zeta^+$ component as opposed to $\cO(y^4)$, cf.~e.g.~(5)-(8) in Ref.~\cite{Strominger:2008dp}. Thus there are transformations that preserve the new set of boundary conditions (\ref{BC}) but not the Brown--Henneaux set of boundary conditions \eqref{eq:false2}. These new transformations must still be considered pure gauge because of their rapid fall-off near the boundary.

Thus we end up with the following situation. The suitable boundary conditions to encompass the logarithmic primary are given by (\ref{BC}) rather than by \eqref{eq:false2}. These are preserved by more gauge transformations than the Brown--Henneaux conditions, but exhibit the same asymptotic symmetries. Since the isometry algebra of AdS$_3$ is part of the transformations that preserve (\ref{BC}), and since the descendants are produced by acting with this algebra, we automatically demonstrated that all descendants of $h^{\rm new}$ fulfill (\ref{BC}).

\section{Boundary stress tensor and asymptotic symmetry generators}

It is also important that all metrics fulfilling (\ref{BC}) have well defined generators of the asymptotic symmetries. This can be shown as follows. We compute the boundary stress tensor along the lines of Ref.~\cite{Grumiller:2008qz} and find a generalization of the Kraus-Larsen result \cite{Kraus:2005zm}:\footnote{Actually, the result given in the first two versions of Ref.~\cite{Grumiller:2008qz} was incorrect due to sign errors. The first appearance of the correct $T_{ij}$ was in Ref.~\cite{Skenderis}. The first correct calculation of the charge $Q$ below was in Refs.~\cite{Henneaux,MSS}. See also our erratum Ref.~\cite{erratum}.}
\begin{align}
T_{++} &= \frac{1}{4\pi G\,\ell}\,h^{\rm new}_{++}\sim \cO(1)\\
T_{--} &= -\frac{1}{8\pi G\,\ell}\,y\partial_y h^{\rm new}_{--}\sim \cO(1)\\
T_{+-} &= 0. 
\end{align}
The off-diagonal contribution $T_{+-}$ vanishes after imposing constraints from the equations of motion. The generators of the asymptotic symmetry group become
\eq{
Q[\zeta]=\frac{1}{4\pi G\,\ell}\, \int dx^+ \,\big(h^{\rm new}_{++}\epsilon^+ - \frac12\,y\partial_y h_{--}^{\rm new}\epsilon^-\big) \sim \cO(1).
}{eq:angelinajolie}
Since no divergences arise the generators \eqref{eq:angelinajolie} are well-defined.

Thus, we conclude that there are indeed consistent boundary conditions \eqref{BC} that go beyond Brown--Henneaux and that allow for the logarithmic primary and all its descendants. Because of the analysis in Section 4 of Ref.~\cite{Grumiller:2008qz} this result might have been anticipated: there it was shown that the logarithmic primary is consistent with the requirement of spacetime being asymptotically AdS and that the ensuing boundary stress tensor is finite, traceless and conserved.

We close by noting that there are other examples when the Brown--Henneaux boundary conditions need to be 
weakened logarithmically to encompass physically interesting solutions~\cite{Henneaux:2004zi}. The boundary conditions of Ref.~\cite{Henneaux:2004zi} are not identical to the ones considered in the present note.


\section*{Acknowledgment}

We thank Stanley Deser, Matthias Gaberdiel, Gaston Giribet, Thomas Hartman, Roman Jackiw, Matthew Kleban, Alex Maloney, Massimo Porrati, Wei Song and Andy Strominger for discussion.

This work is supported in part by funds provided by the U.S. Department of Energy (DoE) under the cooperative research agreement DEFG02-05ER41360. DG is supported by the project MC-OIF 021421 of the European Commission under the Sixth EU Framework Programme for Research and Technological Development (FP6).


\end{document}